%% file: Main.tex
\documentclass[aps,pra,twocolumn,superscriptaddress,showpacs,longbibliography]{revtex4-2}

\usepackage{amsmath}
\usepackage{amssymb}
\usepackage{babel}
\usepackage{bm}
\usepackage{braket}
\usepackage{multirow}
\usepackage{comment}
\usepackage[normalem]{ulem}
\usepackage{here} 

\usepackage{graphicx}
\usepackage{color}
\usepackage[colorlinks=true, linkcolor=magenta,citecolor=blue,urlcolor=magenta]{hyperref}
\DeclareGraphicsExtensions{.pdf}	

\newcommand{\TrS}[1]{\text{Tr}_{\text{S}} [#1]}

\newcommand{\Ex}[1]{\langle #1 \rangle}

\def\F{\mathrm F}

\def\S{\mathrm S}

\def\C{\mathcal C}
\def\D{\mathcal D}
\def\L{\mathcal L}
\def\Ld{{\mathcal L}^{\dagger}}

\def\c{\mathrm c}
\def\s{\mathrm s}

\def\ph{\mathrm{ph}}
\def\tot{\mathrm{tot}}

\def\ds{\delta_{\s}}

\def\Real{\mathrm{Re}}

\def\SS{\mathrm{ss}}

\def\ii{\mathrm i}
\def\dd{\mathrm d}

\def\oA{\hat{A}}
\def\oAd{\hat{A}^{\dagger}}

\def\oX{\hat{X}}
\def\oY{\hat{Y}}

\def\oH{\hat{H}}
\def\oHS{\hat{H}_{\mathrm S}}

\def\osgm{\hat{\sigma}}

\def\usgm{\breve{\sigma}}

\def\oac{\hat{a}_{\c}}
\def\oacd{\hat{a}_{\c}^{\dagger}}

\def\uac{\breve{a}_{\c}}
\def\uacd{\breve{a}_{\c}^{\dagger}}

\def\oa{\hat{a}}

\def\orho{\hat{\rho}}

\def\oOd{\hat{O}^{\dagger}}

\def\uO{\breve{O}}
\def\uOd{\breve{O}^{\dagger}}

\begin{document}
\title{Time-resolved physical spectrum in cavity quantum electrodynamics}

\author{Makoto Yamaguchi}
\altaffiliation{E-mail: makoto.yamaguchi@tokai.ac.jp}
\affiliation{Department of Physics, Tokai University, 4-1-1 Kitakaname, Hiratsuka, Kanagawa 259-1292, Japan}
\author{Alexey Lyasota}
\affiliation{Laboratory of Physics of Nanostructures, Institute of Physics, Ecole Polytechnique F\'{e}d\'{e}rale de Lausanne (EPFL), Lausanne, Switzerland}
\affiliation{Centre of Excellence for Quantum Computation and Communication Technology, School of Physics, University of New South Wales, Sydney, New South Wales 2052, Australia}
\author{Tatsuro Yuge}
\affiliation{Department of Physics, Shizuoka University, Shizuoka 422-8529, Japan}
\author{Yasutomo Ota}
\affiliation{Department of Applied Physics and Physico-Informatics, Keio University, Yokohama 223-8522, Japan}

\date{\today}

\begin{abstract}
The time-resolved physical spectrum of luminescence is theoretically studied for a standard cavity quantum electrodynamics system.
In contrast to the power spectrum for the steady state, the correlation functions up to the present time are crucial for the construction of the time-resolved spectrum, while the correlations with future quantities are inaccessible because of the causality, i.e., the future quantities cannot be measured until the future comes.
We find that this causality plays a key role to understand the time-resolved spectrum, in which the Rabi doublet can never be seen during the time of the first peak of the Rabi oscillation.
Furthermore, the causality can influence on the transient magnitude of the Rabi doublet in some situations.
We also study the dynamics of the Fano anti-resonance, where the difference from the Rabi doublet can be highlighted.
\end{abstract}
\maketitle

\section{Introduction}
The luminescence spectrum is one of the key subjects in quantum optics to study the fundamental characteristics of quantum emitters~\cite{Mandel95, Scully97, Carmichael99, Khitrova06}.
The accessibility to the spectrum is experimentally straightforward, the techniques of its calculation are theoretically well-established, and the consistency between experiment and theory is sufficient, especially when the system of interest is in a steady state.
As a result, in the regime of cavity quantum electrodynamics (cQED), a number of quantum optical effects,
    such as the off-resonant cavity feeding~\cite{Hennessy07, Winger09, Yamaguchi08-2, Ota09, Yamaguchi09, Suffczynski09, Ates09, Jarlov16, Yamaguchi12},
    the incoherent pumping effects~\cite{Laussy08,Laussy09,Valle09},
    the resonance fluorescence~\cite{Valle10, Roy11, Koshino11}
    and the Fano resonance~\cite{Barclay09,Ota15},
have been revealed in the last decades by the steady-state power spectrum, formally described as
\begin{align}
    S_{\SS}(\nu) =  \lim_{t \rightarrow \infty} \int^{\infty}_{0}\dd \tau \Real \sum_{\mu,\mu'} \chi_{\mu,\mu'}\Ex{\uOd_{\mu}(t)\uO_{\mu'}(t+\tau)}_{0} e^{\ii\nu\tau}.
    \label{eq:Sss}
\end{align}
Here, $\uO_{\mu}(t)$ is the system operator linked to the radiation field in the Heisenberg picture, $\Ex{\uOd_{\mu}(t)\uO_{\mu'}(t+\tau)}_{0} = \TrS{\uOd_{\mu}(t)\uO_{\mu'}(t+\tau) \orho_0}$ is the correlation function with $\orho_0$ being the initial density operator, and $\chi_{\mu',\mu}$ denotes a coefficient relevant to each emission channel.
In such a treatment, one advantage to consider the steady state is the equivalence between the correlation functions, $\Ex{\uOd_{\mu}(t)\uO_{\mu'}(t+\tau)}_{0} = \Ex{\uOd_{\mu}(t-\tau)\uO_{\mu'}(t)}_{0}$, which means that the correlation from the future toward the present is statistically equivalent to that from the present toward the past by assuming $\tau>0$ with the present time $t$.
This nature allows us to conveniently forget the causality that the correlations with future quantities cannot be measured until the future comes, and the power spectrum can readily be evaluated through the quantum regression theorem (QRT)~\cite{Mandel95,Scully97,Carmichael99}.

In principle, however, the most fundamental is the time-resolved spectrum because it can naturally resolve dynamical effects~\cite{Ishida13, Peng14}, while extra care is needed for its description, called the physical spectrum~\cite{Eberly77, Mirza15, Ancheyta19, Mejia21}, to avoid artifactual results.
Earlier studies indeed pointed out that the time-resolved spectrum can exhibit distinctive features, e.g., in the fluorescence of a coherently driven atom~\cite{Eberly80, Florjaczyk85, Ho88, Moelbjerg12, Ancheyta18}.
Pump-probe ultrafast spectroscopy has also shown the importance of time-resolved measurements~\cite{Reiter17, KlaBen21, Kasprzak10, Englund12, Majumdar12, Kim13, Bose14, Colman16}.
Nevertheless, the inherent nature of the time-resolved spectrum has not been studied in the context of the causality.
In fact, to our knowledge, no attention has been paid thus far to the impact of the causality.
Moreover, theoretical predictions have increasing importance especially in recent experiments of cQED~\cite{Suffczynski09, Kasprzak10, Englund12, Majumdar12, Kim13, Bose14, Colman16, Lyasota17, Kuruma18} with the Fano interference effect~\cite{Yamaguchi08, Barclay09, Ota15, Bekele19, Yamaguchi21} although there remain few theoretical studies to directly investigate the time-resolved physical spectrum (TRPS) for such systems.

In this paper, we theoretically study the TRPS of luminescence for an initially excited two-level system (TLS) interacting with a single mode cavity.
In its construction, the correlation functions up to the present time are crucial, while the correlations with future quantities are unavailable because of the causality.
We find that this causality gives a key insight to understand the TRPS in the strong coupling regime, where the Rabi doublet can never be seen during the time of the first peak of the Rabi oscillation.
Furthermore, we demonstrate that the magnitude of the Rabi doublet can vary with time in some situations.
These features are in contrast to the common view in the steady-state power spectrum
and significant, e.g., for the shaping of the temporal waveform of single photons and the switching of the Rabi oscillations~\cite{Jin14, Johne15, Pellegrino18, Casabone21}, because the spectral nature of single photons is essential for quantum information applications.
We also study the dynamics of the Fano anti-resonance, where the difference from the Rabi doublet can be highlighted.

\section{Formulations}
Based on the quantum master equation (QME) approach, in the Schr\"odinger picture with $\hbar = 1$, the system density operator, $\orho_{t}$, is evolved by
\begin{align}
    \frac{\dd}{\dd t} \orho_{t} = \L\orho_{t} \equiv -\ii [\oHS,\orho_t] + \D \orho_{t},
    \label{eq:QME}
\end{align}
where $t$ is the present time, and $\D$ is given by
\begin{align}
    \D =&\frac{\gamma}{2}\D_{\osgm_{+},\osgm_{-}}+\frac{\kappa}{2}\D_{\oacd,\oac}+\gamma_{\ph}\D_{\osgm_{+}\osgm_{-},\osgm_{+}\osgm_{-}} \nonumber\\
    &+\frac{\gamma_{\F}}{2}\D_{\oac,\osgm_{+}}+\frac{\gamma^{*}_{\F}}{2}\D_{\osgm_{-},\oacd},
    \label{eq:Dissipator}
\end{align}
with $\D_{\oX,\oY}\orho_{t}=2\oY\orho_{t}\oX-\oX\oY\orho_{t}-\orho_{t}\oX\oY$.
The system Hamiltonian, $\oH_{\S}=\frac{1}{2}\omega_{21}\osgm_{z} + \omega_{\c}\oacd\oac + (g\osgm_{+}\oac+g^{*}\oacd\osgm_{-})$, models the coupling between the TLS
and the cavity by a coupling constant $g=|g|e^{\ii \pi/2}$, where $\oacd$ and $\oac$ are the creation and annihilation operators of the cavity photons,
$\osgm_{\pm}$ are the raising and lowering operators of the TLS,
and $\osgm_{z}=\osgm_{+}\osgm_{-}-\osgm_{-}\osgm_{+}$.
In Eq.~\eqref{eq:Dissipator}, $\gamma$ and $\kappa$ denote the emission rates of the TLS and the cavity, respectively, while $\gamma_{\ph}$ is the pure dephasing rate of the TLS.
The last two terms in Eq.~\eqref{eq:Dissipator} describe the Fano interference between the two emission processes, where $\gamma_{\F}=e^{\ii \theta}\sqrt{\eta\gamma\kappa}$~\cite{Yamaguchi21}.
$\eta$ ($0 \le \eta \le 1$) is the degree of overlap between the two radiation patterns, while $\theta$ is the phase difference between the two processes.

For the construction of the TRPS, it is essential to remember that the QME is originally derived from
the total Hamiltonian including the environment although the QME is a convenient starting point in our analysis.
By considering the Heisenberg equations for the radiation modes~\cite{Scully97,Yamaguchi21}, then, one can arrive at the following expression of the TRPS~\cite{Eberly77},
\begin{align}
    &S(\nu,t,\ds) =  \int^{\infty}_{0}\dd t' \ds e^{-\ds t'} \int^{\infty}_{0}\dd \tau  \nonumber\\
    &\quad \Real \sum_{\mu,\mu'} \chi_{\mu,\mu'}\Ex{\uOd_{\mu}(t-t'-\tau)\uO_{\mu'}(t-t')}_{0} e^{\left(\ii\nu-\frac{\ds}{2}\right)\tau},
    \label{eq:S}
\end{align}
where $\ds$ is the spectral resolution of the spectrometer and $\mu, \mu' \in \{\sigma, a\}$ with $\uO_{\sigma}(t)=\usgm_{-}(t)$ and $\uO_{a}(t)=\uac(t)$.
In the derivation, we have integrated the spectrum over all directions, and as a result, the coefficients $\chi_{\mu',\mu}$ can be determined as $\{\chi_{\sigma,\sigma}, \chi_{a,a}, \chi_{\sigma,a}, \chi_{a,\sigma} \}=\{\gamma/\pi, \kappa/\pi, \gamma_{\F}/\pi, \gamma^{*}_{\F}/\pi\}$.
Here, $S(\nu,t,\ds)$ can recover $S_{\SS}(\nu)$ in the limit of $t \to \infty$, apart from the spectral resolution.
However, in comparison with Eq.~\eqref{eq:Sss}, we can explicitly find the causality in Eq.~\eqref{eq:S}, where the correlations are inherently with the past quantities, instead of the future quantities.
For transient dynamics, $\Ex{\uOd_{\mu}(s-\tau)\uO_{\mu'}(s)}_{0}$ cannot be replaced by $\Ex{\uOd_{\mu}(s)\uO_{\mu'}(s+\tau)}_{0}$ because $\Ex{\uOd_{\mu}(s-\tau)\uO_{\mu'}(s)}_{0}$ depends not only on the time difference $\tau$ but also on the absolute time $s$ ($=t-t'$ in Eq.~\eqref{eq:S}).
This indicates that the causality can have significant impact on the spectrum in general; this is one of the major differences, e.g., from Ref.~\cite{Ishida13}, where the time-resolved spectrum was also discussed.
It is further worth noting that the spectral resolution, $\ds$, is inevitably included in Eq.~\eqref{eq:S} and its influence cannot be separated from the TRPS in general.
This is another important point although the effect of the resolution is often underestimated.

To calculate the TRPS, however, one difficulty lies in its numerical cost due to the double convolution integral, which must be calculated repeatedly for different $\nu$ and $t$.
In our view, this is one reason for the scarcity of theoretical research of the TRPS in cQED systems.
For the reduction of the numerical cost, based on the QRT, we write the correlation as
\begin{align}
    \Ex{\uOd_{\mu}(s-\tau)\uO_{\mu'}(s)}_{0}&= \TrS{(e^{\Ld\tau}\oOd_{\mu'})^{\dagger} \orho_{s-\tau}\oOd_{\mu}} \nonumber \\
    & = \sum_{i}C_{\mu',i}(\tau)\Ex{\oOd_{\mu}\oA_{i}}_{s-\tau},
    \label{eq:MQRT}
\end{align}
for $s \ge \tau \ge 0$,
where $\Ld$ is the adjoint superoperator of $\L$,
$\Ex{\oX}_{t} \equiv \TrS{\oX\orho_{t}}$,
$C_{\mu',i}(\tau) \equiv \TrS{(e^{\Ld\tau}\oOd_{\mu'})^{\dagger}\oAd_{i}}$, and
$\{\oA_{i}\}$ is a complete set of system operators in the Liouville space.
The first equality is helpful in general because $e^{\Ld\tau}\oOd_{\mu'}$ can be computed separately from the evolution of the density operator, $\orho_{s-\tau}$.
The second equality is further advantageous when $C_{\mu',i}(\tau)$ can be obtained analytically.
By employing the latter approach, we have performed one of the integrals in Eq.~\eqref{eq:S} analytically.
We also note that $\Ex{\uOd_{\mu}(s-\tau)\uO_{\mu'}(s)}_{0}=0$ either for $s-\tau < 0$ or for $s < 0$ by definition~\cite{Note1}.
As a result, we can successfully analyze the TRPS; see also Appendix~\ref{App:A} for further details.

\begin{figure*}
\centering
\includegraphics[width=.98\textwidth]{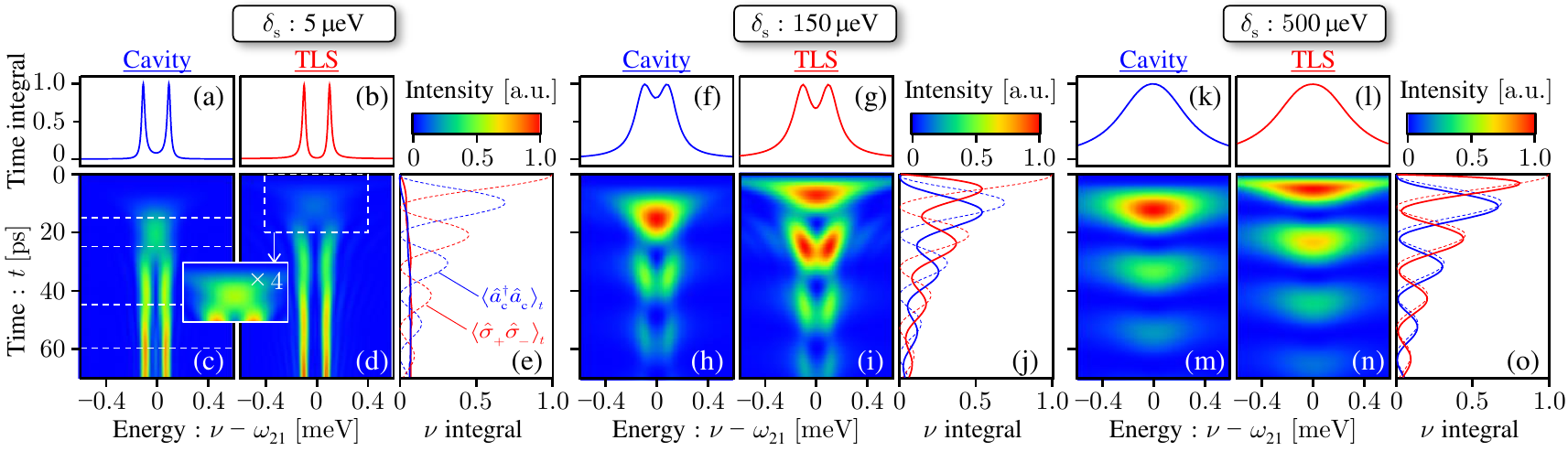}
\caption{
    Calculations for $\gamma_{\ph} = 0$ and $\eta =0$ with $|g|=100~\mu$eV.
    The spectral resolutions are $\ds=5~\mu$eV [left; (a)--(e)], $\ds=150~\mu$eV [middle; (f)--(j)], and $\ds=500~\mu$eV [right; (k)--(o)].
    In the left panels, the time-integrated spectra are shown for the cavity (a) and the TLS (b).
    The spectra are then time-resolved in (c) and (d), which are in turn integrated over energy in (e).
    The intensities are normalized to 1 in (a)--(d) for comparison.
    In (e), the results for the cavity (blue solid curve) and the TLS (red solid curve) are normalized by $\kappa$ and $\gamma$, respectively, to compare with $\Ex{\oacd\oac}_{t}$ and $\Ex{\osgm_{+}\osgm_{-}}_{t}$ (dashed curves).
    The middle and right panels are arranged in the same way.
    The intensity of the area surrounded by the dashed line in (d) is four times increased in the inset.
    $\kappa=50~\mu$eV and $\gamma=0.05~\mu$eV.
}
\label{fig:TRS}
\end{figure*}

\section{Numerical results}
Figure~\ref{fig:TRS} shows typical numerical results for different spectral resolutions with $\gamma_{\ph}=0$ and $\eta=0$.
The TLS is initially excited and other parameters are assumed $\kappa=50~\mu$eV, $\gamma=0.05~\mu$eV, and $|g|=100~\mu$eV under the resonant condition.
Since our setting, $|g| \gtrsim \kappa + \gamma$, indicates the strong-coupling regime, the Rabi doublets should be well resolved when $\ds \ll  |g|=100~\mu$eV.
As a result, for $\ds=5~\mu$eV, we can see the typical Rabi doublets in the time-integrated spectra for the cavity and the TLS [Figs.~\ref{fig:TRS}(a) and (b)].
However, a single-peak structure can be found in the early stage of the TRPS for the cavity [Fig.~\ref{fig:TRS}(c)].
A similar structure can also be seen for the TLS [Fig.~\ref{fig:TRS}(d) and the inset] although its appearance is earlier than for the cavity.
These features are in contrast to the ordinary understanding of the Rabi doublets.
We also note that the dynamics in the TRPS are largely different from the Rabi oscillations of $\Ex{\oacd\oac}_{t}$ and $\Ex{\osgm_{+}\osgm_{-}}_{t}$, as seen in Fig.~\ref{fig:TRS}(e).

By increasing $\ds$ to $150~\mu$eV (comparable to $|g|$), however, oscillatory behaviors come out in Fig.~\ref{fig:TRS}(j), while the Rabi doublets become blurred in Figs.~\ref{fig:TRS}(f) and (g).
This is due to the uncertainty between energy and time.
Hence, by further increasing $\ds$ up to $500~\mu$eV, the energy integrated intensities recover the Rabi oscillations almost completely in Fig.~\ref{fig:TRS}(o), while the Rabi doublets cannot be resolved any more in Figs.~\ref{fig:TRS}(k) and (l).
In fact, in Eq.~\eqref{eq:S}, $\ds e^{-\ds t'}$ gives the time uncertainty as a response function of the spectrometer, while $e^{-\ds\tau/2}$ results in the energy uncertainty as a resolution of the spectrometer.
It is thus obvious that $\ds$ plays an essential role for the consistent description of the physical spectrum.
As a result, we note that the initial rising time of the TRPS intensities cannot be faster than $\ds^{-1}$ in Fig.~\ref{fig:TRS}.

\begin{figure}
\centering
\includegraphics[width=0.48\textwidth]{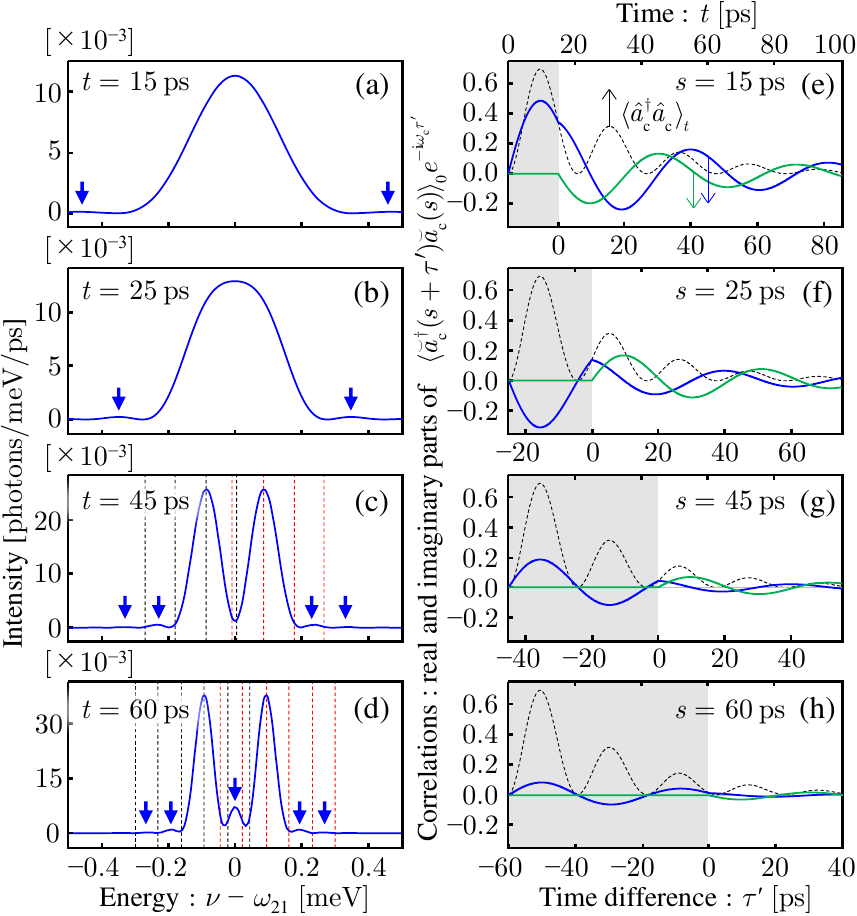}
\caption{
    The cavity transient spectra for $\ds=5~\mu$eV (left panels) and the cavity correlations (right panels).
    In the left panels, the arrows are guides for the eye pointing to the satellite peaks.
    In (c) and (d), the black (red) dashed lines indicate the energy separations from the left (right) main peak by $2\pi n/t$ for several values of the integer $n$.
    In the right panels, the blue (green) solid curves indicate the real (imaginary) parts of the correlations.
    The dashed curves are $\Ex{\oacd\oac}_{t}$ for comparison.
    The correlations vanish for $\tau'<-s$, and therefore, the $\tau'$-axes are drawn from $\tau'=-s$.
    The gray shaded areas indicate the time windows of the causality, $-s<\tau'<0$.
}
\label{fig:ex+crr}
\end{figure}

Beyond such complementarity, we can notice that the single peaks observed in Figs.~\ref{fig:TRS}(c) and (d) are more highlighted in Figs.~\ref{fig:TRS}(h) and (i) and finally form the first peaks of the Rabi oscillations in Figs.~\ref{fig:TRS}(m) and (n).
This indicates that the mechanism of the single peaks is particularly related to the first peaks of the Rabi oscillations.
This is consistent with the earlier appearance of the single peak for the TLS than for the cavity.
However, it is still unclear why such a peculiar role is given only for the first peak.
For a detailed discussion, the cavity spectra for $\ds=5~\mu$eV are separately plotted at different times in Figs.~\ref{fig:ex+crr}(a)--(d), corresponding to the white dashed lines in Fig.~\ref{fig:TRS}(c).
As a result, we can further find that there are weak satellite series around the main peaks, which are absent in the time-integrated spectrum.

To clarify the underlying physics, it is essential to directly analyze the correlation functions.
For this purpose, in Figs.~\ref{fig:ex+crr}(e)--(h), we plot the cavity correlations $\Ex{\uacd(s+\tau')\uac(s)}_{0}e^{-\ii\omega_{\c}\tau'}$ with $\tau' \equiv -\tau$, at absolute times $s$ coincident with the individual times $t$ of Figs.~\ref{fig:ex+crr}(a)--(d).
These correlations at $s=t$ ($t'=0$) give the major contributions in the $t'$ integral of Eq.~\eqref{eq:S} when Figs.~\ref{fig:ex+crr}(a)--(d) are calculated, respectively.
In Figs.~\ref{fig:ex+crr}(e)--(h), we can see that the correlations show oscillatory dynamics at half the Rabi frequency both for $\tau'<0$ and $\tau'>0$.
However, we note that the causality is satisfied only for $\tau'<0$ when calculating the individual spectra [Fig.~\ref{fig:ex+crr}(a)--(d)].
Hence, the correlations become relevant only within the time windows of $-s<\tau'<0$ (the gray shaded areas), as they vanish for $\tau'<-s$~\cite{Note1}.
In fact, this is the reason for the appearances of the satellite series in Figs.~\ref{fig:ex+crr}(a)--(d).
Based on the Fourier analysis, such a time window yields the satellite peaks separated by $2\pi/t$, as indicated by the dashed lines in Figs.~\ref{fig:ex+crr}(c) and (d).
This understanding is consistent with our results that such a phenomenon can be still observed even without the cavity (Appendix~\ref{App:B}).
Furthermore, we can also notice that the Rabi doublets can be observed in Figs.~\ref{fig:ex+crr}(a)--(d) only when the correlations can show their oscillations at least for one period during the time windows in Figs.~\ref{fig:ex+crr}(e)--(h).
In Figs.~\ref{fig:ex+crr}(g) and (h), the correlations oscillate more than one period during the time windows, and therefore, we can see the Rabi doublets in Figs.~\ref{fig:ex+crr}(c) and (d).
In contrast, in Figs.~\ref{fig:ex+crr}(e) and (f), the correlations cannot show their oscillations within the time windows of the causality.
This means that the spectrometer cannot know the oscillating behaviors of the correlations in advance, before it really detects the oscillations at least over one period.
As a result, the single peaks are obtained in Figs.~\ref{fig:ex+crr}(a) and (b).
This is the very reason why the single peaks appear in Figs.~\ref{fig:TRS}(c), (d), (h), and (i).
Since the correlations oscillate at half the Rabi frequency, we can conclude that the Rabi doublets can never be seen during the times of the first peaks of the Rabi oscillations, either for the cavity or for the TLS.
Such unique impact of the causality is not present in the steady-state spectrum.
These results are important by considering that spectral features are, in general, essential for single photons to elicit their fundamental performance in quantum information applications.

We note, however, that the importance of the causality is not limited to the appearances of the satellite series and/or the disappearances of the Rabi doublets.
To illustrate this, we here find that the separations of the Rabi doublets are changed with time, especially for $\ds=150~\mu$eV [Figs.~\ref{fig:TRS}(h) and (i)], where the separations are gradually decreased during the individual Rabi cycles.
This is in contrast to the intuition that the separations are determined only by the magnitude of the coupling constant, $|g|$.
In fact, however, this characteristics can be understood simply from a one-sided Fourier transform, $F_{\phi}(\nu) \equiv \int^{\infty}_{0}\dd \tau \cos(|g|\tau+\phi)e^{\ii\nu\tau-\frac{\ds}{2}\tau}$.
For $\ds$ comparable to $|g|$, $\sin(|g|\tau)e^{-\frac{\ds}{2}\tau}$ is close to a single pulse, while $\cos(|g|\tau)e^{-\frac{\ds}{2}\tau}$ remains more likely to be oscillatory.
Hence, the separation of the doublet in $|F_{\phi}(\nu)|^2$ can be changed by its phase, $\phi$, as well as its frequency, $|g|$ (see also Appendix~\ref{App:C}).
The transient Rabi doublets can then be understood essentially in the same manner.
The important points here are that the corresponding phase is determined by $\Ex{\uOd_{\mu}(s+\tau')\uO_{\mu'}(s)}_{0}$ not for $\tau'>0$ but for $\tau'<0$ due to the causality, and that these two phases are different, as seen in Figs.~\ref{fig:ex+crr}(e)--(h), for example.
These facts mean that the transient Rabi doublets are also strongly influenced by the causality, even if the frequencies of the correlations are the same between $\tau'>0$ and $\tau'<0$.
For example, the dynamics of the cavity Rabi doublet could become very different from Fig.~\ref{fig:TRS}(h) if $\Ex{\uacd(s-\tau)\uac(s)}_{0}$ was replaced by $\Ex{\uacd(s)\uac(s+\tau)}_{0}$ in Eq.~\eqref{eq:S} by neglecting the causality.

\begin{figure}
\centering
\includegraphics[width=0.48\textwidth]{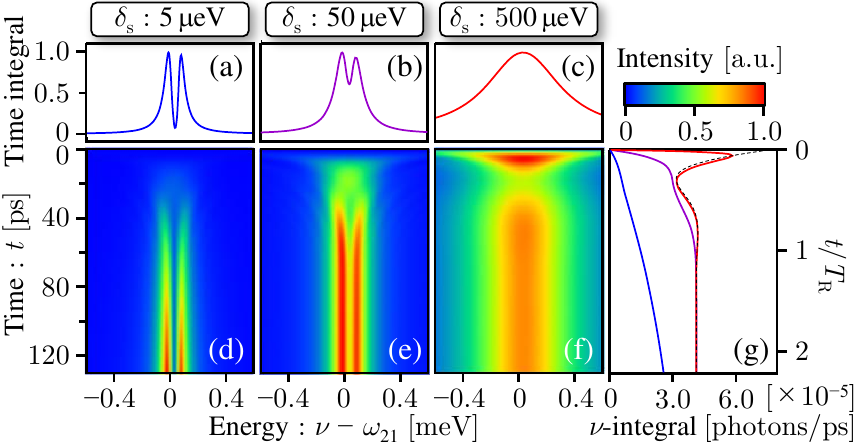}
\caption{
    Calculations for $\gamma_{\ph} = 30~\mu$eV and $\eta =1$ with $|g|=1~\mu$eV, where $\omega_{21}-\omega_{\c}=-70~\mu$eV.
    The time-integrated spectra are shown in the upper panels (a)--(c), while the time-resolved spectra are shown in the lower panels (d)--(f), respectively.
    In (g), the time-resolved spectra are integrated over energy for $\ds=5~\mu$eV (blue), $50~\mu$eV (purple), and $500~\mu$eV (red), where the black dashed curve shows $\kappa\Ex{\oacd\oac}_{t}+\gamma\Ex{\osgm_{+}\osgm_{-}}_{t}+2\Real[\gamma_{\F}\Ex{\osgm_{+}\oac}_{t}]$.
    $T_{\mathrm{R}}=2\pi/\Omega_{\mathrm{R}} \simeq 59~$ps with $\Omega_{\mathrm{R}}=\sqrt{(\omega_{21}-\omega_{\c})^2+4|g|^2}$.
    $\kappa=50~\mu$eV, $\gamma=0.05~\mu$eV, and $\theta=\pi/2$.
}
\label{fig:Fano}
\end{figure}

We have thus clarified the fundamental aspects of the TRPS beyond the conventional view of the Rabi doublet and oscillation.
However, we finally remark that the dynamics of the spectra can largely depend on the inherent mechanism of the doublet in general.
To see this, in Fig.~\ref{fig:Fano}, the Fano interference is studied by setting $\eta =1$ with $\gamma_{\ph} = 30~\mu$eV and $|g|=1~\mu$eV.
In this situation, the Fano anti-resonance yields a doublet in the time-integrated (total) spectrum for $\ds=5~\mu$eV in Fig.~\ref{fig:Fano}(a), while it becomes unclear by increasing $\ds$ in Figs.~\ref{fig:Fano}(b) and (c), as expected.
In contrast to Fig.~\ref{fig:TRS}, however, there is no oscillation relevant to the doublet in Figs.~\ref{fig:Fano}(d)--(g) because the doublet is caused by the destructive interference.
Hence, the presence of the doublet does not necessarily indicate the oscillation.
On the other hand, the single peaks can again be found in the early stages of the TRPS.
This is essentially due to the broadening by the time windows of the causality, but cannot be related to oscillatory behaviors in the time domain.
It follows that the separation of the doublet does not depend on time, even when the doublet is partly blurred by the resolution [Figs.~\ref{fig:Fano}(b) and (e)].
In this context, the time-resolved spectrum is indeed helpful to resolve the dynamical effects.

\section{Conclusions and outlook}
To summarize, we have demonstrated that the TRPS can exhibit distinctive features beyond the steady-state spectrum, where key viewpoints are given by the causality.
Furthermore, the TRPS can also highlight the difference between the Fano anti-resonance and the Rabi doublet.
The approach presented here would have a wide range of applications, such as the shaping of the temporal waveform of single photons and the real-time switching of Rabi oscillations toward quantum information processing~\cite{Jin14, Johne15, Pellegrino18, Casabone21}.
Especially, it provides a useful framework in analyzing the Fano interference from quantum optical systems~\cite{Miroshnichenko10, Limonov17, Litvinenko21, Limonov21}.
In addition, despite our focus on the TRPS, we stress that the significance of the causality is not limited to such a case because of its conceptual generality.
Our results suggest that the pump-probe absorption spectrum~\cite{Reiter17, KlaBen21} and out-of-time-order correlations in photodetection~\cite{Blocher19}, for example, might be non-trivial.
Our findings provide fundamental insight into the time-domain spectroscopy in cQED systems.

\begin{acknowledgments}
M.Y. greatly appreciates fruitful discussion with Prof. Susumu Noda and Dr. Takashi Asano at Kyoto University, Japan, where part of this work was done during the author's doctoral studies (2007--2010).
This work was supported by Japan Society for the Promotion of Science KAKENHI Grant No. JP18K03454.
\end{acknowledgments}

\appendix
\begin{widetext}
\section{The correlation functions and the physical spectrum}\label{App:A}
We here explain our approach to calculate the correlation functions [Eq.~\eqref{eq:MQRT}] and the physical spectrum [Eq.~\eqref{eq:S}].
For this purpose, we consider other correlation functions in the form, $\Ex{\uOd_{\mu}(s)\uO_{\mu'}(s+\tau)}_{0}$, for $\tau \ge 0$ and $s \ge 0$.
In the same manner as $\Ex{\uOd_{\mu}(s-\tau)\uO_{\mu'}(s)}_{0}$ in Eq.~\eqref{eq:MQRT}, $\Ex{\uOd_{\mu}(s)\uO_{\mu'}(s+\tau)}_{0}$ can be written as
\begin{align}
    \Ex{\uOd_{\mu}(s)\uO_{\mu'}(s+\tau)}_{0}
    = \TrS{(e^{\Ld\tau}\oOd_{\mu'})^{\dagger} \orho_{s}\oOd_{\mu}}
    = \sum_{i}C_{\mu',i}(\tau)\Ex{\oOd_{\mu}\oA_{i}}_{s},
    \label{seq:QRT}
\end{align}
where $\{\oA_{i}\}$ is a complete set of system operators in the Liouville space.
The adjoint superoperator $\mathcal{A}^{\dagger}$ for any superoperator $\mathcal{A}$ is defined in such a way that $\TrS{(\mathcal{A}^{\dagger}\oX)^{\dagger}\oY} = \TrS{\oX^{\dagger}\mathcal{A}\oY}$ for arbitrary system operators $\oX$ and $\oY$.
As illustrated in the main text, however, we note that $\Ex{\uOd_{\mu}(s)\uO_{\mu'}(s+\tau)}_{0} \ne \Ex{\uOd_{\mu}(s-\tau)\uO_{\mu'}(s)}_{0}$ for transient dynamics.
Nevertheless, the advantage of this type of correlation function is that the quantum regression theorem (QRT)~\cite{Mandel95,Scully97,Carmichael99} can directly be applied to analyze $\frac{\dd}{\dd \tau}\Ex{\uOd_{\mu}(s)\uO_{\mu'}(s+\tau)}_{0}$ for $\tau \ge 0$.
In contrast, $\frac{\dd}{\dd \tau}\Ex{\uOd_{\mu}(s-\tau)\uO_{\mu'}(s)}_{0}$ for $\tau \ge 0$ is outside the range of its direct application.
Hence, in this paper, we use the fact that $C_{\mu',i}(\tau)$ in Eq.~\eqref{seq:QRT} is the same as in Eq.~\eqref{eq:MQRT}.
Specifically, we can calculate $\Ex{\uOd_{\mu}(s-\tau)\uO_{\mu'}(s)}_{0}$ in the following two steps:
\begin{enumerate}
    \item We first consider $\Ex{\uOd_{\mu}(s)\uO_{\mu'}(s+\tau)}_{0}$ and analytically derive $C_{\mu',i}(\tau)$ by applying the QRT.
    \item We calculate $\Ex{\uOd_{\mu}(s-\tau)\uO_{\mu'}(s)}_{0}$ by using the obtained $C_{\mu',i}(\tau)$ in Eq.~\eqref{eq:MQRT}.
\end{enumerate}
As a result of the first step, for $\mu, \mu' \in \{\sigma, a\}$ with $\uO_{\sigma}(t)=\usgm_{-}(t)$ and $\uO_{a}(t)=\uac(t)$, we obtain
\begin{align}
    &C_{\sigma,1}(\tau) = \frac{1}{\gamma_{+}-\gamma_{-}}
    \left\{ \left( \gamma_{+} + \ii\omega_{\c} + \frac{\kappa}{2} \right) e^{\gamma_{+}\tau}
     - \left( \gamma_{-} + \ii\omega_{\c} + \frac{\kappa}{2} \right) e^{\gamma_{-}\tau} \right\},
    \label{seq:C11}\\
    &C_{a,2}(\tau) = \frac{1}{\gamma_{+}-\gamma_{-}}
    \left\{ \left( \gamma_{+} + \ii\omega_{21} + \frac{\gamma}{2} + \gamma_{\ph} \right) e^{\gamma_{+}\tau}
     - \left( \gamma_{-} + \ii\omega_{21} + \frac{\gamma}{2} + \gamma_{\ph} \right) e^{\gamma_{-}\tau} \right\},
    \label{seq:C22}\\
    &C_{\sigma,2}(\tau) = \frac{-\ii g_{-}}{\gamma_{+}-\gamma_{-}}(e^{\gamma_{+}\tau}-e^{\gamma_{-}\tau}),
    \label{seq:C12}\\
    &C_{a,1}(\tau) = \frac{-\ii g_{+}^{*}}{\gamma_{+}-\gamma_{-}}(e^{\gamma_{+}\tau}-e^{\gamma_{-}\tau}),
    \label{seq:C21}
\end{align}
by assuming $\oA_{1}=\osgm_{-}$ and $\oA_{2}=\oa_{\c}$, where $\gamma_{\pm}$ is defined by
\begin{align}
    \gamma_{\pm} \equiv -\frac{1}{2}(\Gamma_{\tot}+\ii(\omega_{21}+\omega_{\c}))
        \pm \frac{1}{2}\sqrt{\left( \frac{\kappa-\gamma}{2} - \gamma_{\ph} + \ii\omega_{\c,21} \right)^2 - 4g^{*}_{+}g_{-}}
\end{align}
with $\omega_{\c,21} \equiv \omega_{\c}-\omega_{21}$, $g_{\pm} \equiv g \pm \ii \frac{\gamma_{\F}}{2}$ and $\Gamma_{\tot} \equiv \frac{\gamma+\kappa}{2}+\gamma_{\ph}$.
In the derivation, we have used the fact that the bases required to describe the density operator are limited because, in our situation, the TLS is always in the ground state when a photon is inside the cavity.
For the same reason, only $\oA_{1}=\osgm_{-}$ and $\oA_{2}=\oa_{\c}$ are considered in Eqs.~\eqref{eq:MQRT} and \eqref{seq:QRT}.
The second step is then straightforward to perform.
In Figs.~\ref{fig:ex+crr}(e)--(h), the correlations for $\tau'<0$ are obtained by Eq.~\eqref{eq:MQRT}, for example.

For the time-resolved physical spectrum (TRPS) [Eq.~\eqref{eq:S}], these analytic results [Eqs.~\eqref{seq:C11}--\eqref{seq:C21}] are advantageous to reduce the numerical cost.
By inserting Eq.~\eqref{eq:MQRT} into Eq.~\eqref{eq:S}, we can rewrite $S(\nu,t,\ds)$ as
\begin{align}
    &S(\nu,t,\ds) = \Real \sum_{\mu,\mu',i}\chi_{\mu,\mu'} \int^{t}_{0}\dd s' \Ex{\oOd_{\mu}\oA_{i}}_{s'}\C_{\mu',i}(\nu,\ds,t-s'),
    \label{seq:S}\\
    &\C_{\mu',i}(\nu,s'',\ds) \equiv \ds \int^{s''}_{0}\dd\tau C_{\mu',i}(\tau) e^{\left(\ii\nu+\frac{\ds}{2}\right)\tau-\ds s''},
    \label{seq:Cint}
\end{align}
where we have changed the integral variables with using $\Ex{\uOd_{\mu}(s-\tau)\uO_{\mu'}(s)}_{0}=0$ either for $s-\tau < 0$ or for $s < 0$.
Therefore, by using Eqs.~\eqref{seq:C11}--\eqref{seq:C21}, the integration of Eq.~\eqref{seq:Cint} can be performed as
\begin{align}
    &\C_{\sigma,1}(\nu,s'',\ds) = \frac{1}{\gamma_{+}-\gamma_{-}}
    \left\{ \left( \gamma_{+} + \ii\omega_{\c} + \frac{\kappa}{2} \right) \C_{+}(\nu,s'',\ds)
     - \left( \gamma_{-} + \ii\omega_{\c} + \frac{\kappa}{2} \right) \C_{-}(\nu,s'',\ds) \right\},
    \label{seq:C11int}\\
    &\C_{a,2}(\nu,s'',\ds) = \frac{1}{\gamma_{+}-\gamma_{-}}
    \left\{ \left( \gamma_{+} + \ii\omega_{21} + \frac{\gamma}{2} + \gamma_{\ph} \right) \C_{+}(\nu,s'',\ds)
     - \left( \gamma_{-} + \ii\omega_{21} + \frac{\gamma}{2} + \gamma_{\ph} \right) \C_{-}(\nu,s'',\ds) \right\},
    \label{seq:C22int}\\
    &\C_{\sigma,2}(\nu,s'',\ds) = \frac{-\ii g_{-}}{\gamma_{+}-\gamma_{-}}(\C_{+}(\nu,s'',\ds)-\C_{-}(\nu,s'',\ds)),
    \label{seq:C12int}\\
    &\C_{a,1}(\nu,s'',\ds) = \frac{-\ii g_{+}^{*}}{\gamma_{+}-\gamma_{-}}(\C_{+}(\nu,s'',\ds)-\C_{-}(\nu,s'',\ds)),
    \label{seq:C21int}
\end{align}
with
\begin{align}
    \C_{\pm}(\nu,s'',\ds) \equiv \ds \frac{e^{\left( \ii\nu+\gamma_{\pm}-\frac{\ds}{2} \right)s''} - e^{-\ds s''} }{\ii\nu+\gamma_{\pm}+\frac{\ds}{2}}.
    \label{seq:Cpmint}
\end{align}
We can thus calculate the TRPS by Eq.~\eqref{seq:S} with Eqs.~\eqref{seq:C11int}--\eqref{seq:Cpmint}.
We note that these results can exactly recover the previous results for the time-integrated spectrum, $\int^{\infty}_{-\infty}\dd t S(\nu,t,\ds)$~\cite{Yamaguchi21}.
On the other hand, the energy-integrated intensity, $\int^{\infty}_{-\infty}\dd \nu S(\nu,t,\ds)$, can also be obtained as
\begin{align}
    \int^{\infty}_{-\infty}\dd \nu S(\nu,t,\ds) = \int^{t}_{0} \dd s''
    \left( \kappa\Ex{\oacd\oac}_{s''} + \gamma\Ex{\osgm_{+}\osgm_{-}}_{s''} + 2\Real[\gamma_{\F}\Ex{\osgm_{+} \oac}_{s''}] \right) \cdot \ds e^{-\ds(t-s'')},
    \label{seq:en-int}
\end{align}
which is the convolution of the relevant quantities with the response function of the spectrometer, $\ds e^{-\ds t}$.
As a result of the response function of the spectrometer, we note that the initial rising times of the TRPS intensities cannot be faster than $\ds^{-1}$ in Fig.~\ref{fig:TRS}.

 \section{Results for the initially excited two-level system without the cavity}\label{App:B}
In the case of the initially excited TLS without the cavity, the TRPS defined for the TLS,
\begin{align}
    &S_{21}(\nu,t,\ds) \equiv \int^{\infty}_{0}\dd t' \ds e^{-\ds t'} \int^{\infty}_{0}\dd \tau
    \Real \frac{\gamma}{\pi}\Ex{\usgm_{+}(t-t'-\tau)\usgm_{-}(t-t')}_{0} e^{\left(\ii\nu-\frac{\ds}{2}\right)\tau},
    \label{seq:S_21_def}
\end{align}
can be obtained analytically, based on the same strategy as used in Appendix~\ref{App:A}.
In the derivation, the TLS correlation, $\Ex{\usgm_{+}(s+\tau')\usgm_{-}(s)}_{0}$, is given by
\begin{align}
    &\Ex{\usgm_{+}(s+\tau')\usgm_{-}(s)}_{0} =
    \begin{cases}
        e^{-\gamma s}e^{\left(\ii\omega_{21}-\frac{\gamma}{2}-\gamma_{\ph}\right)\tau'} & (\tau' \ge 0)\\
        e^{-\gamma s}e^{\left(\ii\omega_{21}-\frac{\gamma}{2}+\gamma_{\ph}\right)\tau'} & (-s \le \tau' < 0)\\
        0 & (\tau' < -s)
    \end{cases},
    \label{seq:TLScrr}
\end{align}
with $s \ge 0$.
As a result, $S_{21}(\nu,t,\ds)$ for $t \ge 0$ is obtained as
\begin{align}
    &S_{21}(\nu,t,\ds) = \frac{\gamma}{\pi}\Real
    \left[
        \frac{ \ds e^{-\gamma t} }
             { \ii(\nu-\omega_{21})-\frac{\gamma}{2}-\gamma_{\ph}+\frac{\ds}{2} }
        \left\{
            \frac{ 1 - e^{\left(\ii(\nu-\omega_{21})+\frac{\gamma}{2}-\gamma_{\ph}-\frac{\ds}{2}\right)t}}
            {-\ii(\nu-\omega_{21})-\frac{\gamma}{2}+\gamma_{\ph}+\frac{\ds}{2}}
            +
            \frac{1-e^{(\gamma-\ds)t}}
            {\gamma-\ds}
        \right\}
    \right].
    \label{seq:S_21}
\end{align}
Here, the expression of the TRPS [Eq.~\eqref{seq:S_21}] is somewhat complicated despite the simple situation.
We note, however, that the time-integrated spectrum has a simple Lorentzian line shape
\begin{align}
    &\int^{\infty}_{0} \dd t S_{21}(\nu,t,\ds) = \frac{1}{\pi}
    \frac{\frac{\gamma+\ds}{2}+\gamma_{\ph}}
        {(\nu-\omega_{21})^2+\left(\frac{\gamma+\ds}{2}+\gamma_{\ph} \right)^2}.
    \label{seq:S_21_TI}
\end{align}
On the other hand, the energy-integrated intensity results in a convolution between $\gamma\Ex{\osgm_{+}\osgm_{-}}_{t}=\gamma e^{-\gamma t}$ ($t \ge 0$) and the response function of the spectrometer, $\ds e^{-\ds t}$,
\begin{align}
    &\int^{\infty}_{-\infty} \dd \nu S_{21}(\nu,t,\ds)
    = \int^{t}_{0}\dd s'' \gamma\Ex{\osgm_{+}\osgm_{-}}_{s''} \cdot \ds e^{-\ds (t-s'')},
    \label{seq:S_21_EI}
\end{align}
which does not depend on the pure dephasing rate, $\gamma_{\ph}$.

\begin{figure*}
\centering
\includegraphics[width=.98\textwidth]{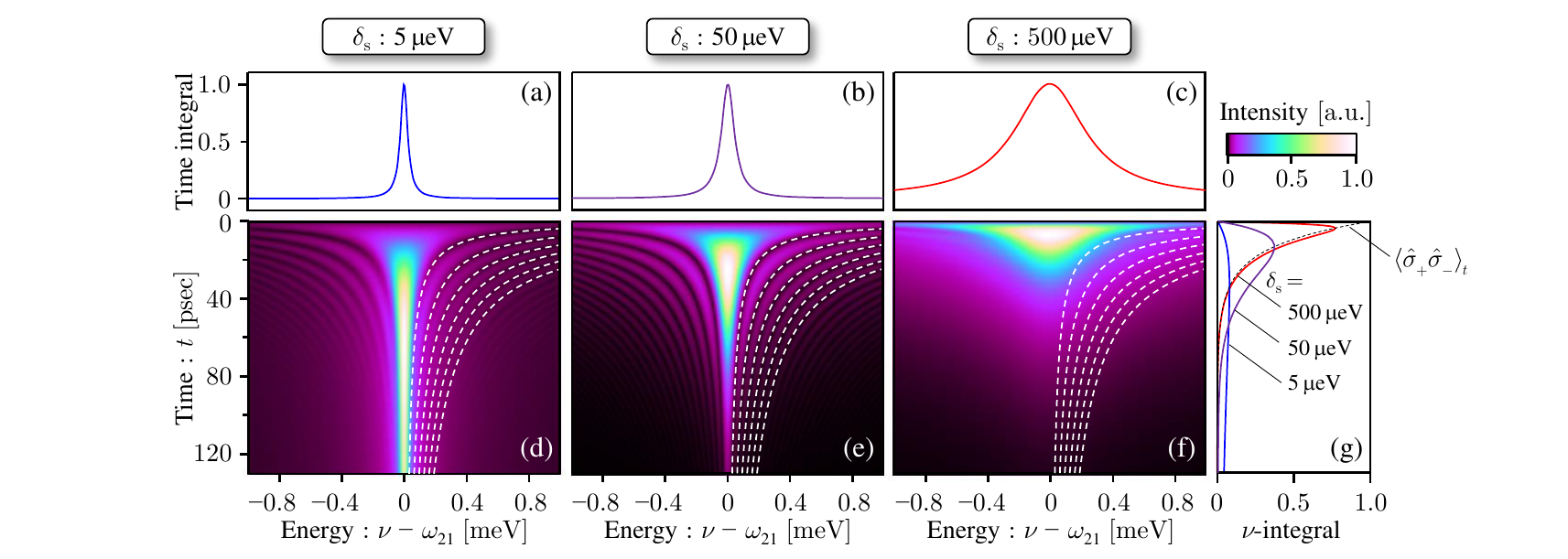}
\caption{
    Numerical results for the TLS with $\gamma=50~\mu$eV and $\gamma_{\ph}=0$.
    For $\ds=5$, $50$ and $500~\mu$eV, the time-integrated spectra, $\int^{\infty}_{0} \dd t S_{21}(\nu,t,\ds)$ [Eq.~\eqref{seq:S_21_TI}], are shown in the upper panels (a)--(c), while the time-resolved spectra, $S_{21}(\nu,t,\ds)$ [Eq.~\eqref{seq:S_21}], are in the lower panels (d)--(f), respectively.
    The intensities are normalized to 1 in (a)--(f).
    In (g), the time-resolved spectra are integrated over energy, $\int^{\infty}_{-\infty} \dd \nu S_{21}(\nu,t,\ds)$ [Eq.~\eqref{seq:S_21_EI}], for $\ds=5~\mu$eV (blue), $50~\mu$eV (purple), and $500~\mu$eV (red), where the intensities are normalized by $\gamma$ to be compared with $\Ex{\osgm_{+}\osgm_{-}}_{t}$ (black dashed curve).
    The white dashed curves in (d)--(f) indicate $\nu-\omega_{21}=2\pi n/t$ for $n=1, 2, \cdots, 6$.
    Results for other values of the integer $n$ are not shown for the sake of visibility.
}
\label{sfig:TLS}
\end{figure*}

Figure~\ref{sfig:TLS} shows typical numerical results for $\gamma=50~\mu$eV and $\gamma_{\ph}=0$.
By increasing $\ds$, we can see that the time-integrated spectra are broadened in Figs.~\ref{sfig:TLS}(a)--(c), while the energy-integrated intensities (normalized by $\gamma$) get closer to the population of the excited state, $\Ex{\osgm_{+}\osgm_{-}}_{t}$, in Fig.~\ref{sfig:TLS}(g).
This is due to the energy-time uncertainty.
Furthermore, we can notice that there are satellite series around the main peaks in Fig.~\ref{sfig:TLS}(d) and (e), which are absent in the time-integrated spectra of Figs.~\ref{sfig:TLS}(a) and (b).
The separation of each satellite peak is well captured by $2\pi/t$, as indicated by the white dashed curves.
This is consistent with the explanation in the main text, due to the time window of the causality.
In contrast, the satellite series cannot be seen in Fig.~\ref{sfig:TLS}(f).
This is because the separation of each satellite peak, $2\pi/t$, is much smaller than the broadening effect by $\ds$.
In this context, we can notice that $t \ll \ds^{-1}$ should at least be satisfied for the appearance of the satellite peaks.
For the same reason, we note that there is no satellite series in Figs.~\ref{fig:TRS}(m) and (n), although not described in the main text.

\section{Influence of the phase on the doublet}\label{App:C}
In this appendix, finally, the influence of the phase on the doublet is discussed by using the one-sided Fourier transform, $F_{\phi}(\nu) = \int^{\infty}_{0}\dd \tau \cos(|g|\tau+\phi)e^{\ii\nu\tau-\frac{\ds}{2}\tau}$.
As shown in Figs.~\ref{sfig:F}(a)--(c), we can see that the intensities of $|F_{\phi}(\nu)|^2$ have some similarities to the time-resolved spectra in Fig.~\ref{fig:TRS}.
In particular, we can find that the separation of the doublet depends on the phase $\phi$ when $\ds$ is comparable to $|g|$ [Fig.~\ref{sfig:F}(b)], as indicated by the white dashed curves.
Here, the separation of the doublet is small for $\phi=-\pi/2$ because $\cos(|g|\tau+\phi)e^{-\frac{\ds}{2}\tau}=\sin(|g|\tau)e^{-\frac{\ds}{2}\tau}$ is close to a single pulse.
In contrast, the separation is large for $\phi=0$ because $\cos(|g|\tau+\phi)e^{-\frac{\ds}{2}\tau}=\cos(|g|\tau)e^{-\frac{\ds}{2}\tau}$ remains more likely to be oscillatory.
We can thus intuitively understand that $|F_{\phi}(\nu)|^2$ can depend on the phase, $\phi$, from the time-domain behaviors.
In contrast, in the frequency domain, we can notice that the two doublet peaks are influenced by each other when the broadening by $\ds$ is comparable to $|g|$ because $F_{\phi}(\nu)$ can be written as
\begin{align}
    F_{\phi}(\nu)=e^{\ii \phi}\frac{\ds}{(\nu+|g|)^2+\ds^2} + e^{-\ii \phi}\frac{\ds}{(\nu-|g|)^2+\ds^2}.
    \label{seq:F}
\end{align}
Hence, it can again be found that the phase, $\phi$, plays an important role in the profile of $|F_{\phi}(\nu)|^2$.

\begin{figure*}
\centering
\includegraphics[width=.98\textwidth]{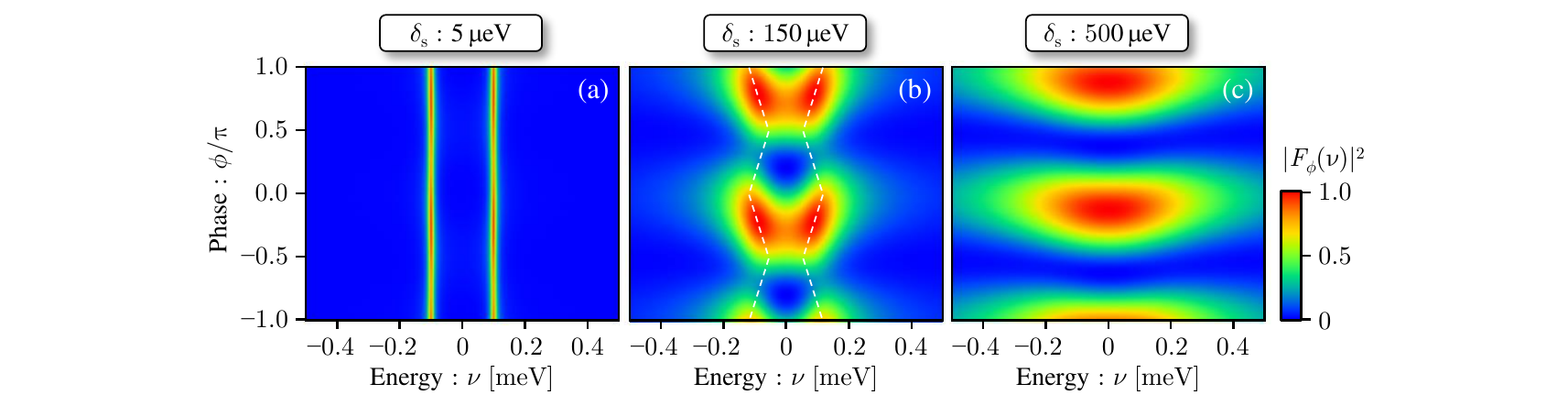}
\caption{
    The phase dependence of $|F_{\phi}(\nu)|^2$ for $|g|=100~\mu$eV.
    The spectral resolution, $\ds$, is assumed $\ds = 5~\mu$eV (a), $150~\mu$eV (b), and $500~\mu$eV (c), where the intensities are normalized to 1 for comparison.
    In (b), the white dashed curves are guides for the eye, indicating the phase dependence of the doublet.
}
\label{sfig:F}
\end{figure*}
\end{widetext}

\footnotetext{
    This is also a kind of the causality because any quantities are consequential only after the initial time $t=0$.
}

\newpage
\input{ref.bbl}

\end{document}

%% file: ref.bbl
%